\documentclass[12pt,preprint]{aastex}

\shorttitle{Space Velocities of Southern Globular Clusters}
\shortauthors{Casetti-Dinescu et al.}

\begin{document}

\title{Space Velocities of Southern Globular Clusters. V. 
A Low Galactic Latitude Sample}

\author{Dana I. Casetti-Dinescu\altaffilmark{1,2}, 
Terrence M. Girard\altaffilmark{1}, David Herrera\altaffilmark{1}, William F. van Altena\altaffilmark{1},
Carlos E. L\'{o}pez\altaffilmark{3} and Danilo J. Castillo\altaffilmark{4}}

\altaffiltext{1}{Astronomy Department, Yale University, P.O. Box 208101,
New Haven, CT 06520-8101, (dana,girard,vanalten@astro.yale.edu, david.herrera@yale.edu)}
\altaffiltext{2}{Astronomical Institute of the Romanian Academy, Str.
Cutitul de Argint 5, RO-75212, Bucharest 28, Romania}
\altaffiltext{3}{Yale Southern Observatory, Universidad Nacional de San Juan,
Avenida Benav\'{i}dez, 8175, Oeste, 5407 Marquesado, San Juan, Argentina (celopez@speedy.com.ar)}
\altaffiltext{4}{Yale Southern Observatory, Barreal, Calingasta, Argentina (daniloeros@hotmail.com)}

\begin{abstract}
We have measured the absolute proper motions of globular clusters
NGC 2808, 3201, 4372, 4833, 5927 and 5986. The proper motions are on the $Hipparcos$ system,
and they are the first determinations ever made for these low Galactic latitude clusters.
The proper motion uncertainties range from 0.3 to 0.5 mas yr$^{-1}$.
The inferred orbits indicate that 1) the single metal rich cluster in our
sample, NGC 5927, dynamically belongs to the thick disk, 
2) the remaining metal poor clusters have rather low-energy orbits
of high eccentricity; among these, there appear to be
two ``pairs'' of dynamically associated clusters, 
3) the most energetic cluster in our sample,
NGC 3201 is on a highly retrograde orbit --- which had already been 
surmised from radial velocity alone --- with an
apocentric distance of 22 kpc, and 4) none of the metal poor clusters 
appear to be associated with the recently detected SDSS streams, or with the
Monoceros structure.

These are the first results of the Southern Proper-Motion Program
(SPM) where the second-epoch observations are 
taken with the recent CCD camera system installed on the double astrograph
at El Leoncito, Argentina.

\end{abstract}


\section{Introduction}

In the past decade considerable progress has been made in understanding
our Galaxy's globular cluster (GC) system. This is in large part due to dedicated programs  
that aim to characterize homogeneously
properties of clusters including chemical abundance patterns, ages, 
horizontal branch (HB) morphology, structural parameters  and orbits. 
Studies have combined the recently aquired various observational databases to search
for correlations that can point to a
realistic formation picture of the globular-cluster
system (see e.g., Carreta 2006, Recio-Blanco et al. 2006,
Pritzl et al. 2005, Mackey \& Gilmore 2004). 
Briefly, the present picture points to an accreted origin for the
outer halo clusters ($r_{GC} \ge 10$ kpc), while 
a combination of dissipational collapse and some accretion can best explain
the current properties of inner GCs. However, details of these
pictures are by no means understood as there appear to be 
a series of ``non-canonical'' observations.
 For instance, there are metal-poor ([Fe/H] $< -1.0$) clusters
with disk-like kinematics (Dinescu et al. 1999b - Paper III and references therein,
Dinescu et al. 2003 - Paper IV), there are metal-rich clusters ([Fe/H] $> -0.8$) with 
unusually blue, extended HBs that appear to reside within the bulge,
(e.g., Rich et al. 1997), and (for the data currently available)
the orbits of the most energetic clusters
appear to be on average more eccentric than those of present-day dwarf spheroidal satellites
(Dinescu et al. 2001). Also, these clusters are not 
dynamically associated with any of the
current streams found in the SDSS (Belokurov et al. 2006, Grillmair \& Dionatos 2006,
see also our Discussion in this study). There is only one exception to this latter
point: cluster Pal 12 and its association with the Sagittarius dwarf galaxy (Sgr;
Dinescu et al. 2000, Martinez-Delgado et al. 2002, Cohen 2004).
Therefore questions such as how much accretion took place in the past to build up the inner halo, 
can it really be traced and separated from the features of a dissipational collapse, and
how different was it
from the ongoing/recent accretion seen in surveys such as SDSS remain to be
explored.

With these questions in mind, we continue our program to determine
absolute proper motions of GCs especially in the inner halo, and thus contribute
new orbit information to the overall picture of
the formation of the Milky Way GC system. 
%This information also contributes to the study of the survival of the GC system
%via calculations of the destruction rates due to gravitational shocks
%(Allen et al. 2006, Baumgart \& Makino 2003, Paper III).
Previous results from our program (Dinescu et al 1997 - Paper I, Dinescu et al 1999a - Paper II,
Paper III and Paper IV) were based on photographic plates alone. The current
results are based on CCD data for the second epoch. A CCD system with two
cameras was mounted in 2003 on the double astrograph at El Leoncito, Argentina, where
our observation are based. Here we show the first astrometric results that make use of the
new CCD system.

In Section 2, we describe the observations and reductions including those of the  CCD system
recently mounted on the astrograph. In Section 3
we describe the proper-motion derivations. Sections 4 
presents the velocity and orbit results, and finally, 
a discussion of the results is given in Section 5.

\section{Observations and Measurements}

This work is part of the continuation of
the Southern Proper Motion Program (SPM, Platais et al. 1998, Girard et al. 1998, 2004),
a survey that aims to produce absolute proper motions and $V, B$ photometry 
for $\sim 100$ million stars in the southern sky, down to $V \sim 17.5$. A recent release
of this program based on photographic material alone, includes 10 million objects
(Girard et al. 2004). The photographic plate material used in this work is 
described in Table 1.

The remaining area for the 
$2^{nd}$ epoch observations is being
mapped with a two-camera CCD system installed on the 51-cm double astrograph 
at Cesco Observatory, El Leoncito,
Argentina. The earliest observations included in the SPM program started in June 2003. 
This system's properties and
performances are briefly described below, while a more detailed description will be
given in a future SPM general-program update. 

The program clusters are a low-latitude sample that supplements our previous work
for 15 high-latitude clusters (Papers I and II). Thus, in the current work, the
proper motions are tied to an inertial reference system via  $Hipparcos$ stars 
rather than galaxies. The first results for this low-latitude sample were
presented for four bulge clusters (l $= 350\arcdeg-360\arcdeg$)
in Paper IV. The current
sample is located in the fourth Galactic quadrant (Table 2). Other limitations on the sample are 
imposed by the SPM first-epoch plate material, i.e., clusters are south of $\delta = -20\arcdeg$
and within $\sim 10$ kpc from the Sun. The novelty of the current work lies in the use of
$2^{nd}$ epoch CCD data that have improved the precision of our proper-motion results
by 1) expanding the baseline from 20-25 years to 30-38 years, and 2) improving
the positional precision of the $2^{nd}$ epoch observations.

\subsection{Photographic Measurements}

The SPM plates were taken with the double astrograph at Cesco Observatory 
in blue (103a-O) and visual (102a-G + OG515 filter) passbands (see Table 1).
The plate scale is 55.1$\arcsec$ mm$^{-1}$, and 
each field covers $6\fdg3 \times 6\fdg3$. SPM plates
contain a 2-hr exposure that reaches to $V\sim 18$ and an offset 2-min exposure.
During both exposures, an objective grating is used, which produces a series of
diffraction images on either side of the central, zero-order images; depending on 
grating orientation, the diffraction images are aligned with the E-W direction in most cases,
and, in a few cases, along the N-S direction (Girard et al. 1998). The multiple sets of
images of bright stars allow us to detect and  model magnitude-dependent systematics that
affect both positions and proper motions. These systematics are 1) present in  
practically all photographic material, 2) mainly due to the nonlinear response of the
photographic emulsion, and 3) require special methods to detect internally  (Girard et al. 1998 and
references therein).

For the SPM plate material, the size of these systematics is 10 to 15 mas over 6 magnitudes;
their modeling and removal using diffraction images has been thoroughly described
in  Girard et al. (1998). This key feature of the 
SPM program to internally correct for magnitude equation 
ensures that bright stars, incluing the astrometric calibrators
($Hipparcos$ stars, $V \sim 9$), and faint stars (e.g., cluster stars,
 $V \ge 14$) are on a system largely 
free of systematics. A comparison 
between the absolute proper motion of NGC 6121 (M4) as determined from the SPM material
and calibrated via $Hipparcos$ stars (Paper II) with that determined from HST data 
and calibrated to 
extragalactic objects (one QSO: Bedin et al. 2003 and eleven galaxies: Kalirai et al. 2004), 
indicates excellent agreement within the quoted uncertainties of $\sim 0.4$ mas yr$^{-1}$.

\begin{table}[htb]
\begin{center}
\caption{SPM Field Characteristics}
\begin{tabular}{ccccl}
%\multicolumn{5}{c}{Table 1. SPM Field Characteristics} \\ \\
\tableline
\\
\multicolumn{1}{c}{NGC} & \multicolumn{1}{c}{Field \#} & \multicolumn{1}{c}{R.A.}
& \multicolumn{1}{c}{Dec.} &
\multicolumn{1}{l}{Plate \# (Epoch)} \\
 & & ($h$ $m$) & ($\arcdeg$) & \\
\tableline
\\
2808  & 096 & $09~20$ & $-65$ & 401BY (1969.04) \\
3201  & 289 & $10~24$ & $-45$ & 095BY (1967.05) \\
4372  & 068 & $12~48$ & $-70$ & 794BY (1972.22) \\
4833  & 068 & $12~48$ & $-70$ & 794BY (1972.22) \\
5927  & 241 & $15~12$ & $-50$ & 295BY (1968.33) 491BY (1969.53) \\
5986  & 362 & $15~36$ & $-40$ & 300BY (1968.33) 477BY (1969.45) \\
\tableline
\end{tabular}
\end{center}
\end{table}

The target clusters and the properties of the SPM fields/plates in which they
were measured are listed in Table 1.
The photographic plates were scanned with the Yale PDS microdensitometer,
in object-by-object mode, with a pixel size of 12.7 microns. On each SPM field, we measure
a pre-selected set of stars (see also Paper IV).
This set
consists
of all $Hipparcos$ and Tycho2 stars (ESA 1997), $\sim 200$ Guide Star Catalog
1.1 (GSC, Lasker et al. 1990) stars,  $\sim 3000$ faint field stars selected from
the USNO-A2.0 catalog (Monet et al. 1998) in the magnitude range 15 to 17, and cluster stars.
For bright stars ($V < 14$) we measure both exposures and diffraction images. 
$Hipparcos$ stars provide the correction to absolute proper motion, while Tycho2 and GSC stars
assure an appropriate magnitude range of various diffraction orders with which to model
magnitude-dependent systematics. The faint stars serve as reference stars, i.e., are used
to map one plate into another, as well as CCD positions into the photographic ones.
Spatially, they are distributed around each $Hipparcos$ 
star, and in a ring around the cluster. This special configuration was chosen
to minimize modeling uncertainties when plate positions are 
transformed into one another (see Paper I). For each SPM field we measure $\sim 100$ $Hipparcos$ 
stars; and there are twenty faint stars surrounding each $Hipparcos$ star, and $\sim 2000$ 
faint stars within the ring surrounding the cluster. The list of cluster stars to be measured on the
plate is determined from a CCD frame (see below) centered on the cluster.
The input positions for these stars are determined from 
the CCD frame and the software package Sextractor (Bertin \& Arnouts 1996).
Cluster stars are selected to cover a region of a few times the half-light radius as
taken from Harris (1996, 2003 update, hereafter H96). The radius of this region  varies between 
$4\arcmin$ and $9\arcmin$ for the various clusters, with the central $1\arcmin - 2\arcmin$ 
being unusable because of crowding.

\begin{figure}
\includegraphics[scale=0.80]{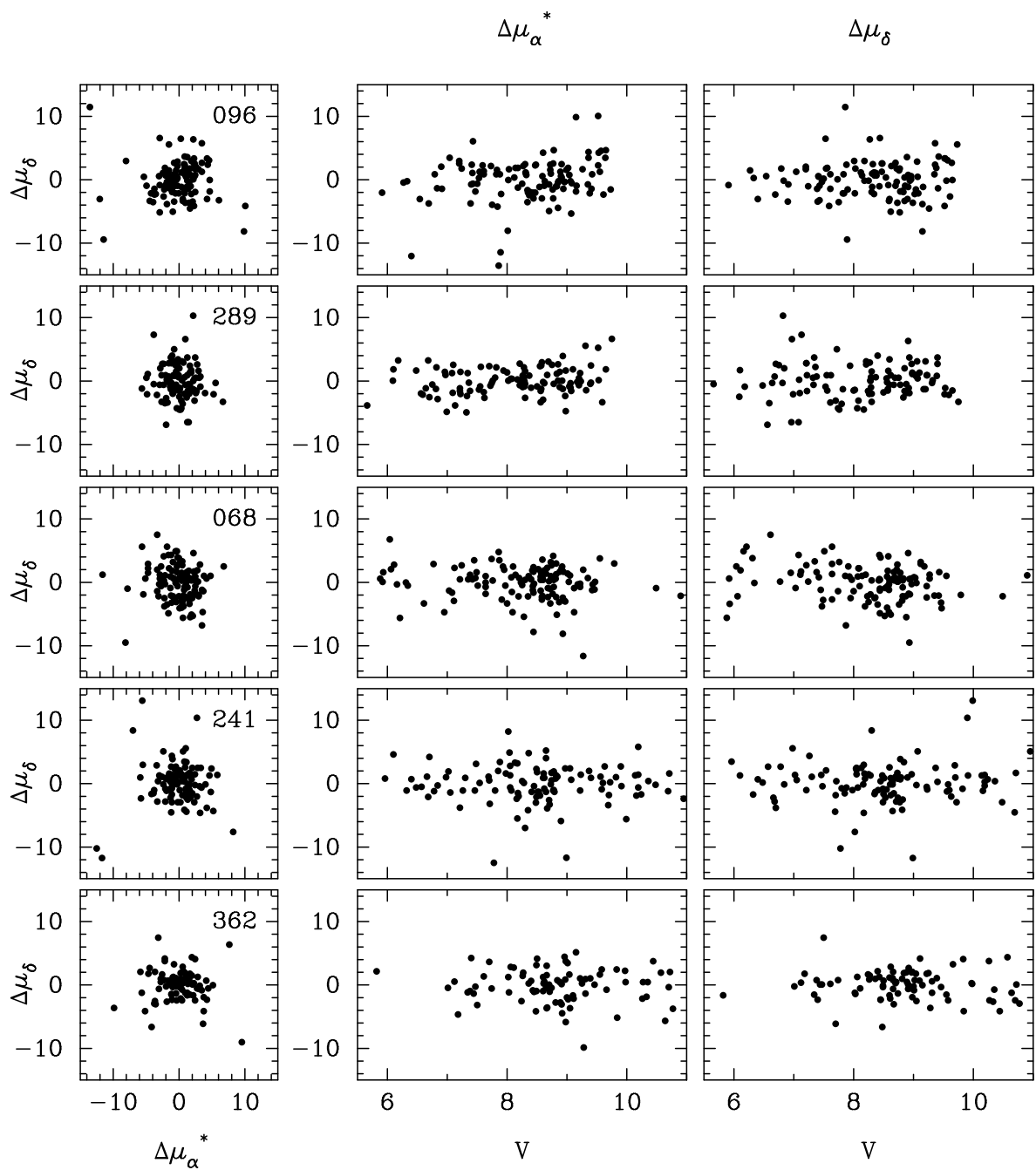}
\caption{Differences between $Hipparcos$ and our proper motions for the five SPM
fields. The average value of the differences, i.e., the correction to absolute proper motion,
has already been applied as on offset such that the proper motion range is the same for all fields.}
\end{figure}

\subsection{CCD Measurements}

The CCD camera system consists of two main cameras, one each in the blue and visual
focal planes of the double astrograph. The PixelVision visual CCD camera has a 4K $\times$ 4K 
unthinned, front-illuminated Loral
chip of total area $0\fdg94 \times 0\fdg94$. The pixel size of $0\farcs83$ is well
matched to our site, where the seeing is typically between $2\arcsec$ and $3\arcsec$, corresponding
to 3-4 pixels per FWHM. With a 2-minute exposure in reasonable seeing conditions, a magnitude 
limit of $V \sim 18$ is reached. The original blue camera was an Apogee AP-8  CCD with a 
1K x 1K SITe back-illuminated chip, covering $0\fdg38 \times 0\fdg38$ at
$1.32\arcsec$/pixel. In May 2005, this camera was 
replaced by an Apogee Alta E42 2K $\times$ 2K camera with a field of view of
$0\fdg42 \times 0\fdg42$ at $0.74\arcsec$/pixel. The observations are taken with the diffraction grating
oriented at $45\arcdeg$. Thus the entire CCD dynamical range covered is 10 magnitudes, 
i.e., 6 inhrently from the CCD plus 4 more provided by the grating.

For the SPM program, the PixelVision camera is the primary astrometry instrument, and $V$-band
photometer, while the Apogee cameras provide $B$ photometry and possibly astrometry over a fraction of the
survey. 
For the cluster work presented here, we make use only of the PixelVision astrometric data.
The planned two-fold overlap coverage with CCD frames requires some 90 frames over an isolated
$6\fdg3 \times 6\fdg3$ SPM field. In addition, six or more frames are taken centered
on each program cluster: each set of three frames has exposures of 30, 60 and 120 sec.
The CCD pre-processing pipeline includes calibrations using biases and flats for the PixelVision frames.
Detections and aperture photometry
are derived with SExtractor software. The positions from SExtractor, which are
intensity-weighted centroids, are then used as initial positions in our refined centering
routine that fits a 2D elliptical Gaussian to the image profile. These later positions have a
centering precision of $\sim 20$ mas per single image for well-measured objects ($V = 7$ to 15).

Before attempting to derive proper motions based on the PixelVision
CCD positions, two separate pre-corrections are performed.
First, a correction for the optical field angle distortion (OFAD) of the
field of view is determined and applied.
Second, positions from the grating-order images must be placed on a
common system with those of the deeper, central-order images.
This latter step is necessary to provide a reliable means of linking
the faint cluster members with the bright $Hipparcos$ stars that will be
used to determine the correction to absolute proper motions.

The OFAD of the PixelVision frames, to the extent that it is constant
and stable, can be determined by averaging the residuals between our CCD
positions and those of an external catalog, over many frames.
General quadratic polynomial transformations between UCAC2 (Zacharias et al. 2000)
coordinates
and the (central-order) positions from the PixelVision frames provides
residuals adequate for this task.
As the telescope pointings used to create the UCAC2 catalog  are distributed randomly
across the PixelVision field of view, any resulting systematic
patterns revealed can be attributed to the combined PixelVision/astrograph
effective OFAD.
We have averaged a minimum of 
$\sim 200,000$ residuals from at least $94$ frames reduced into the UCAC2
using a quadratic transformation, for each SPM field in this study.
For each field, 
an empirical correction mask is derived, on a $21 \times 21$
grid across the 4K $\times$ 4K pixel field.
This mask is applied to the positions from each PixelVision frame, using
bilinear interpolation within the mask grid. 
Typical amplitudes of the position corrections are 10 to 15 mas.

The second crucial step in the processing of PixelVision positions is the
unification of the various diffraction-order coordinate systems.
Given perfect optics and detector, and ignoring the possible effects
of differential color refraction, the average
position of the two first-order image centers should coincide exactly
with the position of the central-order image.
Likewise, the average of the positions of the two second-order images
should also be coincident with the central order image.
An offset can indicate the presence of magnitude equation - the magnitude
dependent bias in stellar image positions often seen in photographic
material but also expected to a lesser degree in CCD data because of
imperfect Charge Transfer Efficiency (CTE).
With the first-order images being effectively four magnitudes fainter than
the corresponding central-order image, the presence of any
magnitude-dependent bias would directly lead to a non-zero offset.
In fact, these positional offsets are used to determine and then
correct the magnitude equation present in the SPM photographic plates,
(see Girard et al. 1998).

In the case of the PixelVision data, our hope was to use the offsets between
central and diffraction-grating images to measure any CTE-induced
magnitude equation and correct for it.
Indeed, substantial offsets on the order of 0.05 pixels ($0.04\arcsec$) are seen,
however these do not appear to be CTE-related, i.e., caused by inherent
magnitude equation.
The offsets between the average of the first-order images 
relative to the central image, $\Delta_{10}$, vary from 0.0 to 0.1 pixels.
These offsets are well-correlated with the coresponding offsets of the average 
of the second-order image pair relative to that of the central image, 
$\Delta_{20}$, within any given frame.
Empirically, we find $\Delta_{20} = 2.0 \Delta_{10}$.
This is most certainly {\it not} the behavior to be expected if the offsets were
due to magnitude equation.
The second-order images are only marginally fainter than the first-order
images, while both are substantially fainter than the central order.
The magnitude differences between grating order and central order are
4.00 and 4.66 for the first and second orders, respectively.
Obviously, the observed positional offsets, $\Delta_{10}$ and $\Delta_{20}$, 
are {\it not} proportional to the magnitude offsets.

A search for possible dependencies between the size of a particular frame's
offsets and hour angle, grating orientation angle, and seeing have all
failed to reveal any underlying cause of the observed offsets.
Noting that the ratio of the first and second-order offsets, 2.0, matches
the ratio of the actual separation of these diffraction images on the frame,
we have decided to interpret the offsets as geometric in nature.
That is, the bias in position is postulated to be proportional to separation 
from the central order image.
Thus, we have adopted a scheme for transformation from grating-order coordinate
system to central-order system that is a uniform
$\Delta X_{10}$, $\Delta Y_{10}$ to be applied to the positions of all 
first-order image pairs, and
$\Delta X_{20}$, $\Delta Y_{20}$ that is applied to the second-order
positions.
The values of $\Delta X_{10}$, $\Delta Y_{10}$, $\Delta X_{20}$, and
$\Delta Y_{20}$ are calculated separately for each PixelVision frame,
using probability plots (Hamaker 1978) of the inner $80\%$ of all measureable
grating-pair/central-image triads that also meet conservative photometric 
criteria to ensure the exclusion of saturated central-order images.

Further details of our analysis that led to the development of
both the positional correction mask and the grating-order offsets that
are applied to the PixelVision positions will accompany the publication of 
the next SPM proper-motion catalog expected later this year.
For the present study, comparisons of the final proper motions
determined using separately the first and second-order images alone will
serve as a check on the latter, and less certain, of these two corrections.

All CCD observations were taken between 2003 and 2005, thus ensuring a baseline of
30 to 38 years (see Table 2).

\section{Proper Motions}

After all precorrections are applied to the photographic plate positions
(i.e., refraction, magnitude systematic-correction, etc., see Girard et al. 1998) and to the
CCD positions (see above), we proceed to calculate proper motions using the
central-plate overlap method (e.g., Girard et al. 1989). All measurements are transformed
into the system of one photographic plate that is used as a master plate. 
We choose a first-epoch visual plate to serve as the master. The remaining $1^{st}$-epoch 
plates are transformed into the master plate 
using a polynomial of up to $4^{th}$ order. The reference stars used for these transformations
have a magnitude range between $ V =10$ and 18; however, the input list is such that
stars with $V \sim 16$ dominate the reference frame. Therefore, it is the central-order 
image that is used in these transformations.
Between $\sim 5000$ and 6800 stars are used
in these plate transformations, 
and the derived positional precision per single measurement
for these stars is between 70 and 85 mas. The central-order transformation from one plate 
into another, is then applied to the remaining orders as well.  

The CCD positions are also mapped into the 
system of the photographic master plate, individually for each CCD frame.
For these transformations, only $2^{nd}$-order polynomials are necessary.
Typically, for each CCD frame, there are $\sim 100$ stars that model this transformation
(see the selection of the input list in Section 2.1), and, as with the plates, only
the central-order image is used. For the $1^{st}$ and $2^{nd}$-order images, we apply the 
transformation defined by the central order. 
For frames that include a cluster, there are between $\sim 500$ and 1800 faint reference
stars that map this transformation;
clusters stars (i.e., those selected within a few half-light radii of the cluster)
 are $not$ used in this transformation. The scatter in this transformation's residuals is due to both
measurement errors and cosmic proper-motion dispersion. 
Proper motions are then determined by treating each image order as an
independent set of positions for both photographic plates and CCD frames.
We have thus the possibility to test determinations based on various image orders,
and plates.
A linear least-squares fit of positions as a function of time gives the proper motion for each object.
Measurements that differ by more than $0\farcs2$ from the best-fit line are considered
outliers, and excluded; the formal
proper-motion uncertainty is given by the scatter about the best-fit line.
These proper motions are thus relative to a reference frame that is dominated
by 16th-magnitude stars. By treating individually each CCD frame, we make the implicit
assumption that the mean motion of the reference system does not vary over a 6$\times$ 6-degree field.
However, this is not necessarily true; in fact we have found linear spatial gradients
across the SPM fields that must be considered when the final cluster 
proper motion is determined (see below).

\subsection{Correction to Absolute Proper Motions}

The correction from a relative reference frame defined by 16th-magnitude stars to 
an inertial reference frame is determined from straight differences between the $Hipparcos$
proper motions and our relative proper motions. In Figure 1 we show these differences for all 
five SPM fields. The left panels show the vector-point diagram (VPD) while the middle and 
right hand panels show the differences in each coordinate as a function of magnitude.
Units of proper motion are mas yr$^{-1}$ throughout the paper, and
$\mu_{\alpha}^* = \mu_{\alpha} cos \delta$.
The offset defined by these differences is the correction to absolute proper motion.
In Fig. 1 we have applied these offsets so that a similar proper-motion range is displayed
for all fields. The offsets are determined using probability plots (Hamaker 1978) trimmed
at $10\%$ at each edge. Other estimators such as simple average and median have also been tested;
they all give consistent results within the estimated uncertainty. 

We obtain a scatter of between 
2 and 3 mas yr$^{-1}$, and consequently,  formal uncertainties in the absolute proper-motion
correction of between 0.2 and 0.3 mas yr$^{-1}$. The proper motions shown are
constructed from all image orders on the
photographic plates and only the $1^{st}$ and $2^{nd}$ order on the CCD frames. 
Central-order images for
$Hipparcos$ stars on the CCD frames are saturated and unusable. For each SPM field, we have
also determined the correction by using separately only blue or yellow plates, or
only CCD $1^{st}$-order or $2^{nd}$-order images, to check for possible systematics. 
The results indicate
that differences between different determinations are within the estimated uncertainties.
The sole variation found was the variation of the proper-motion differences 
across the field. The size of the gradients is $\le 0.01$ mas yr$^{-1}$ mm$^{-1}$ which  
will amount to a significant deviation, when compared to formal uncertainties,
for clusters that lie far from the center of the
spatial distribution of the $Hipparcos$ stars. We have therefore applied adjustments to
the absolute proper-motion correction that account for these gradients, for each cluster. 
We have also verified that these spatial gradients are 
of the magnitude expected for the change in the mean
motion of the reference system across the field. We have used the Besan\c{c}on Galactic model (Robin et
al. 2003, 2004) to predict the mean proper-motion gradient across SPM field 068, and 
confirm the variation seen in our measures.

\subsection{Cluster Proper Motions}

The mean motion of the cluster with respect to field stars is determined 
from the stars measured in the cluster region and trimmed in the color-magnitude diagram
(CMD) for clusters
NGC 2808, NGC 4833, NGC 5927 and NGC 5986. For the remaining clusters, NGC 3201 and
NGC 4372, we have applied a two-component Gaussian fitting procedure for the proper motion distributions, as the 
field contribution is considerable, even after trimmed via the CMD. 
In the cases of NGC 2808, 5927 and 5986, we have used 2MASS (Cutri et al. 2003) J and K photometry to 
select cluster stars in the CMD. For NGC 4833, we have used B and V photometry from the
study of Melbourne et al. (2000), that  covers the entire area of our measured cluster stars
(a circle with $6\farcm8$ radius). In Figures 2 and 3 we show the CMDs and the proper-motion 
distributions for each cluster. The open symbols represent all stars
measured in the cluster area, while the black symbols those selected from the CMD to be likely
cluster members. The left panels
show the CMDs, the middle panels show the VPDs of all the stars in the cluster area. The right 
panels show the VPDs of the stars selected from the corresponding CMD. The histograms show
the marginal distributions of the proper motions. The cross  shows the adopted mean
relative proper motion of the cluster. For clusters NGC 2808, 4833, 5927 and 5986, we have determined
this mean with probability plots trimmed at $10\%$ of CMD-selected stars within a radius of
20 mas yr$^{-1}$ from the approximate centroid of the cluster proper-motion distribution.
Thus stars outside this proper-motion range are considered outliers. The remaining field contribution
after the CMD selection, is assumed to be small and therefore eliminated by the 
$10\%$ cut in the probability-plot determination. 

\begin{figure}
\includegraphics[scale=0.85]{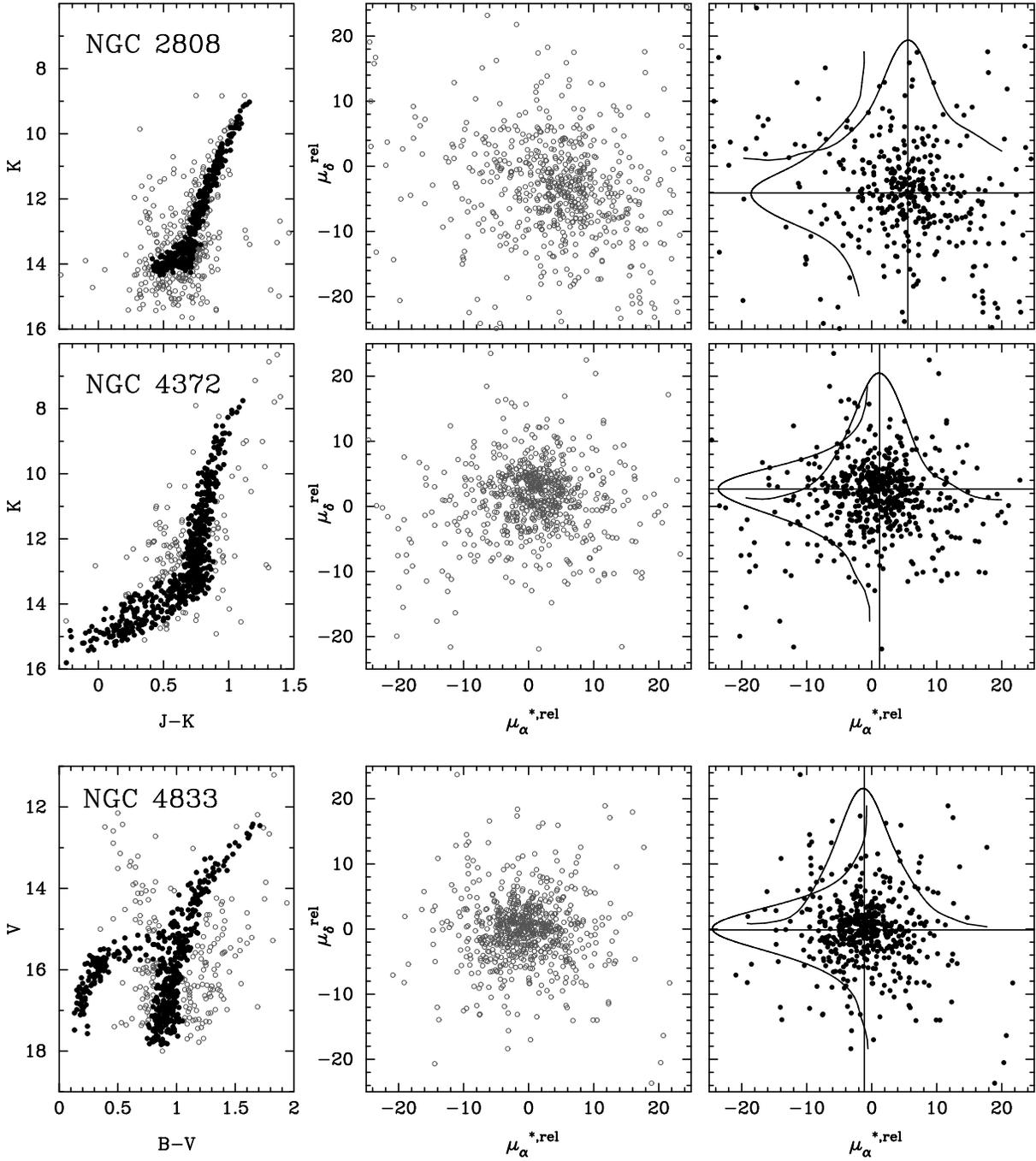}
\caption{CMDs and relative proper-motion distributions for each cluster field. Dark symbols represent
the CMD-selected cluster members. The marginal distributions for CMD-selected members are also shown.
The cross marks the adopted mean relative proper motion of the cluster (see text).}
\end{figure}

\begin{figure}
\includegraphics[scale=0.85]{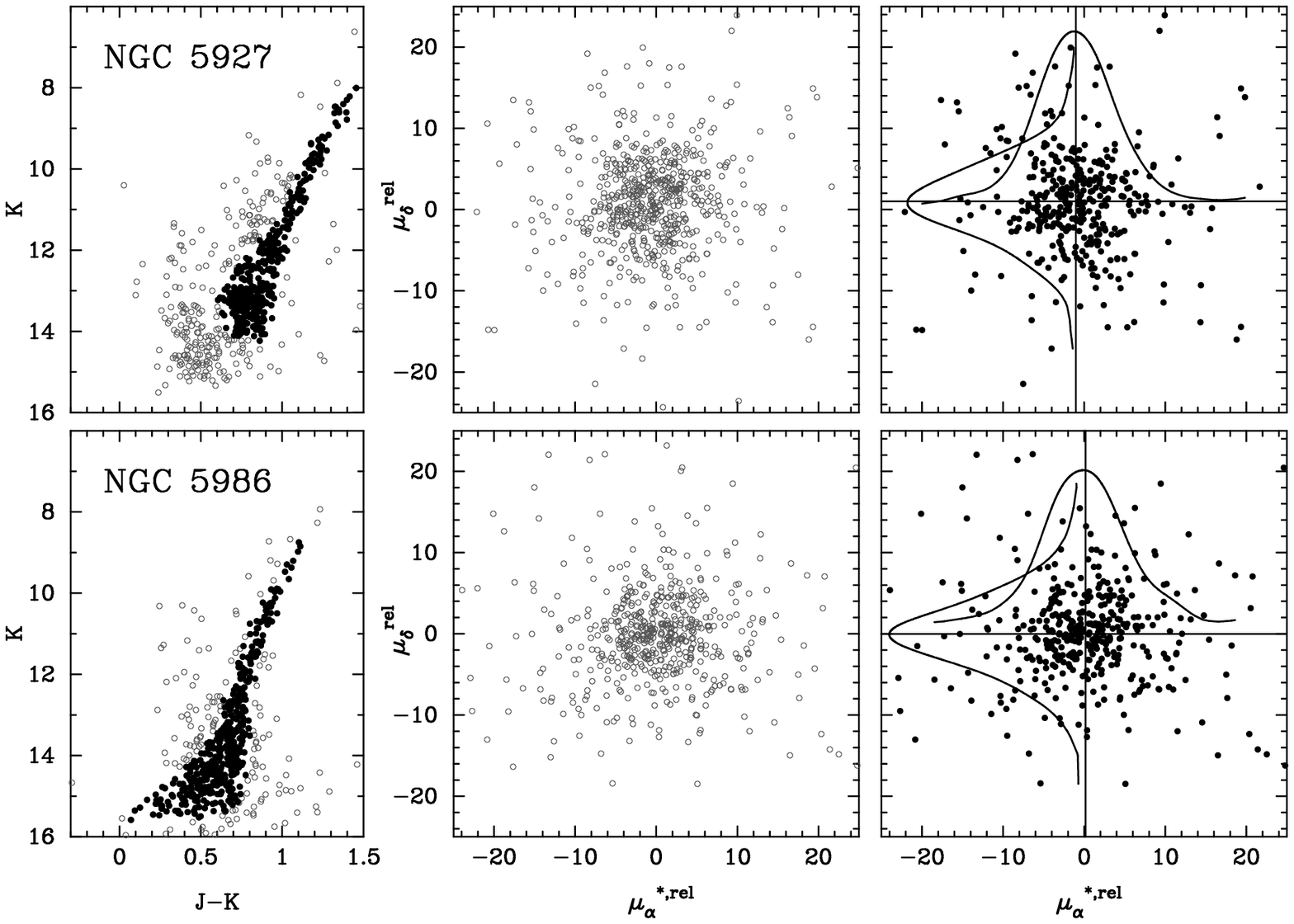}
\caption{Same as Fig. 2, only for clusters NGC 5927, and NGC 5986}
\end{figure}

This was not the case however for
NGC 4372, and NGC 3201. Since the area where cluster stars were measured for these two clusters is
larger than that of the other four clusters,  as they are more extended, 
it is likely that more field contamination contributes
to the cluster sample, even when a CMD selection is applied. That this is the case for NGC 4372,
can be seen in the $\mu_{\delta}$ marginal distribution  (Fig. 2 - second right-hand panel), 
which is visibly skewed. Therefore we chose to fit the proper-motion
distribution of all of the stars in the cluster area (i.e., no selection using the
CMD) with a model consisting of the sum of two Gaussians, one representing
NGC 4372, the other the field. This is done separately for each coordinate, and the 
``observed'' proper-motion distribution is constructed from the data smoothed with 
the individual proper-motion errors. Details of this procedure can be found
for instance in Girard et al. (1989). The mean and width of the fitted Gaussian to  
the cluster sample 
represent the mean proper motion and proper-motion uncertainty of the cluster.

\begin{figure}
\includegraphics[scale=0.85]{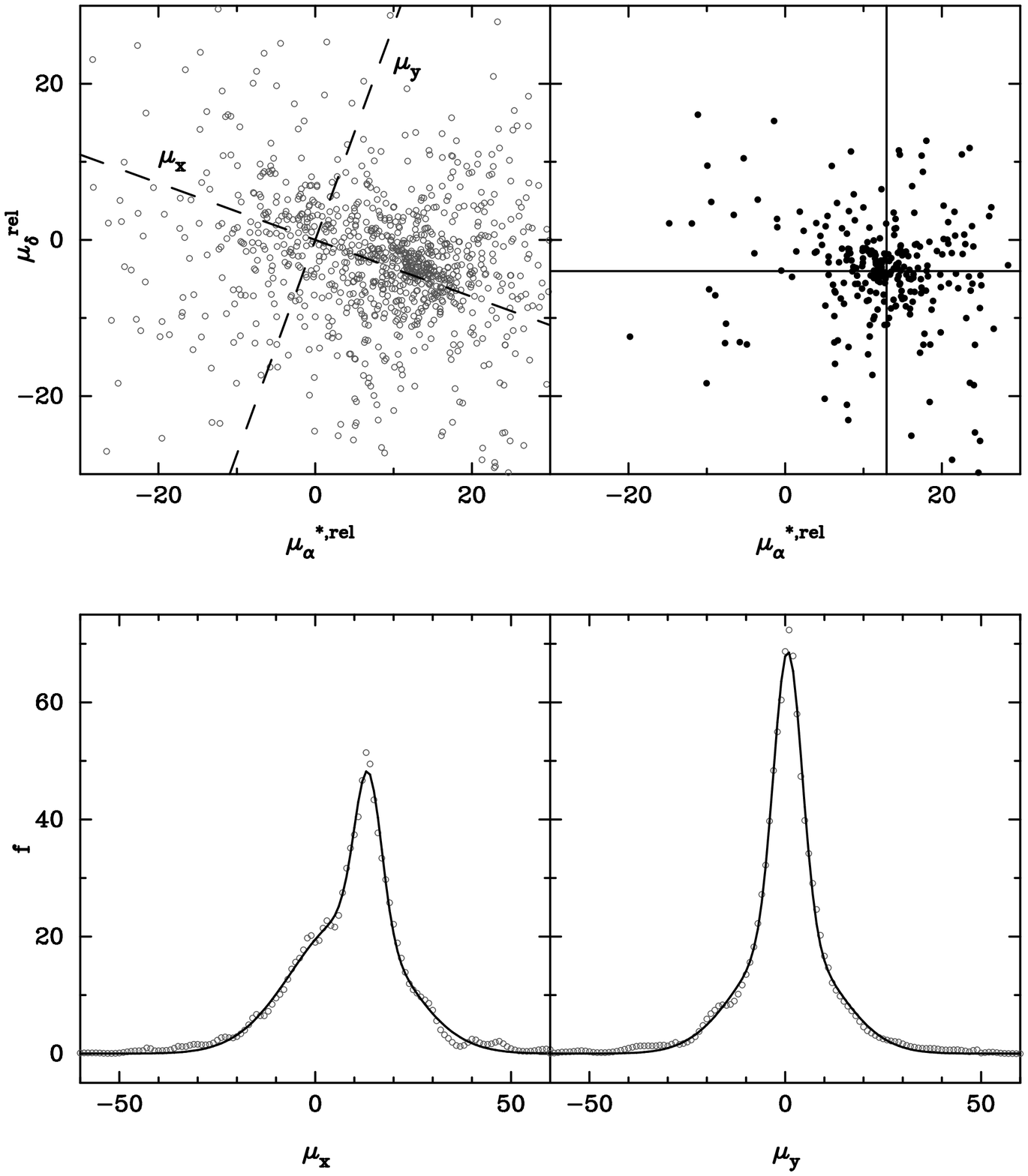}
\caption{The relative proper-motion distribution in the field of NGC 3201 for
all stars (top, left panel), and for radial-velocity cluster members (top, right).
The bottom panels show the marginal distributions along the rotated ($\mu_x, \mu_y$)
system (see text). The continuous line shows the two-component
 Gaussian fit to the data.
The cross in the top-right panel is the mean relative proper motion of the
cluster as determined from the two-component Gaussian fit to the data in the top-left panel.}
\end{figure}

A similar procedure was 
applied to the field of NGC 3201. This is illustrated in Figure 4. The top left panel
of Fig. 4 shows the proper-motion distribution of all stars measured in the cluster area.
The dotted lines show the rotated system in which the Gaussian fit is
made, aligning the $x$ axis with the 
elongated shape of the field proper-motion distribution.
In the top, right panel we show the radial-velocity selected sample of cluster stars.
The radial velocities are from C\^{o}te et al. (1995); with a mean of 494 km s$^{-1}$,
they are thus very distinct from the field radial-velocity distribution. The cross marks 
the mean proper motion as determined from the Gaussian fit. 
This fit is shown in the bottom panels of Fig. 4. The fit is applied to the entire sample 
of stars measured in the cluster area; the open symbols show the observed 
proper-motion distribution, and the black line shows the fit.
We have also calculated the mean motion of the cluster by using only the radial
velocity-selected stars and the probability plot estimator with no trimming.
Only stars laying outside a 20 mas yr$^{-1}$ radius from the proper-motion 
centroid are excluded as measurement outliers. From this determination,
we obtain $(\mu_{\alpha}^*, \mu_{\delta}) = (13.32 \pm 0.35, -3.72 \pm 0.33)$ mas yr$^{-1}$.
The Gaussian fit gives $(\mu_{\alpha}^*, \mu_{\delta}) = (12.93 \pm 0.22, -3.99 \pm 0.18)$ mas yr$^{-1}$.
Since the results agree very well we are convinced that the field contamination
is well modeled and appropriately accounted for 
in the Gaussian fit. Therefore we adopt this value.

The number of cluster stars that enter into each determination varies between 260 and 400 stars.
As in the case of correction to absolute proper motions, we have searched for 
possible systematics by performing separate solutions for the blue and visual plates.
In the case of NGC 4833, since the sample allowed it, we have also looked at the
mean proper motion as determined from the blue, horizontal branch stars, and the red giant
stars. We have found that results agreed within estimated uncertainties in all cases
except for NGC 4372. Here, the blue plate solution proved different from that of the visual 
plate, and with the blue-plate solution showing a large scatter in the cluster star proper motions.
Since NGC 4372 lies in the corner of the SPM field, it is likely that this area of the
first-epoch blue plate is damaged. This was not seen however in the field of NGC 4833,
which lies on the same SPM field (Table 1) but near the center of the plate. 
We have therefore eliminated the blue plate measurements from NGC 4372's proper-motion
determination.

\begin{table}[tbh]
\begin{center}
\caption{Cluster Properties and Absolute Proper Motions}
\begin{tabular}{rrrrrrrrr}
%\multicolumn{9}{c}{Table 2. Cluster Properties and Absolute Proper Motions} \\ \\
\tableline
\\
\multicolumn{1}{c}{NGC} & \multicolumn{1}{c}{l} &  \multicolumn{1}{c}{b} &  \multicolumn{1}{c}{$d_{\odot}$} &
\multicolumn{1}{c}{V$_{rad}$} & \multicolumn{1}{c}{[Fe/H]} & \multicolumn{1}{c}{M$_V$} &
\multicolumn{1}{c}{$\mu_{\alpha} cos \delta$} & \multicolumn{1}{c}{$\mu_{\delta}$} \\
& \multicolumn{1}{c}{($\arcdeg$)} & \multicolumn{1}{c}{($\arcdeg$)} & 
\multicolumn{1}{c}{(kpc)} & \multicolumn{1}{c}{(km s$^{-1}$)} & & & 
\multicolumn{1}{c}{(mas yr$^{-1}$)} & \multicolumn{1}{c}{(mas yr$^{-1}$)} \\
\tableline
\\
2808 & 282.2 & -11.3 & 9.6 & 93.6 & -1.15 & -9.39 & $0.58\pm0.45$ & $2.06\pm0.46$ \\
3201 & 277.2 &   8.6 & 5.0 & 494.0 & -1.58 & -7.46 & $5.28\pm0.32$ & $-0.98\pm0.33$ \\
4372 & 301.0 & -9.9 & 5.8 & 72.3 & -2.09 & -7.77 & $-6.49\pm0.33$ & $3.71\pm0.32$ \\
4833 & 303.6 & -8.0 & 6.5 & 200.2 & -1.80 & -8.16 & $-8.11\pm0.35$ & $-0.96\pm0.34$ \\
5927 & 326.6 & 4.9 & 7.6 & -107.5 & -0.37 & -7.80 & $-5.72\pm0.39$ & $-2.61\pm0.40$ \\
5986 & 337.0 & 13.3 & 10.4 & 88.9 & -1.58 & -8.44 & $-3.81\pm0.45$ & $-2.99\pm0.37$ \\
\tableline
\end{tabular}
\end{center}
\end{table}

The final absolute proper motions are listed in Table 2, along with other cluster 
parameters taken from H96. Our proper-motion uncertainties, which include 
contributions from 
both
the absolute correction and the cluster mean relative motion,
are between  0.3 and 0.5 mas yr$^{-1}$.

\section{Velocities and Orbital Parameters}

Velocities are calculated assuming the solar circle radius $R_{0} = 8.0$ kpc, and
the rotation velocity of the Local Standard of Rest (LSR) $\Theta_{0} = 220$ km s$^{-1}$.
The adopted solar peculiar motion is $(U_{\odot}, V_{\odot}, W_{\odot}) = (-10.00, 5.25, 7.17)$ km s$^{-1}$
(Dehnen \& Binney 1998); $U$ is positive outward from the Galactic center. In Table 3, we list
the current location of each cluster and its velocity components in a cylindrical
coordinate system. Uncertainties in the velocity components include uncertainties in the proper motion,
radial velocity, and an adopted $10\%$ error in the distance; in Table 3, they are the
numbers in parentheses. Also, in Table 3, $z$ is the displacement
 perpendicular to the
Galactic plane, and R$_{GC}$ is the distance from the Galactic center projected onto the Galactic plane.
Of the six clusters, the most straightforward cases of kinematical classification
as inferred solely from the velocities and current
Galactic locations, are NGC 5927 and NGC 3201. NGC 5927, a metal rich cluster (Table 2), has 
kinematics consistent with membership to the thick disk of the Galaxy, while NGC 3201 has a 
strongly retrograde orbit, as originally suspected from its radial
velocity alone (e.g., Gonzalez \& Wallerstein 1998).

\begin{table}[tbh]
\begin{center}
\caption{Galactic Positions and Velocities}
\begin{tabular}{rrrrrr}
%\multicolumn{6}{c}{Table 3. Galactic Positions and Velocities} \\ \\
\tableline 
\\
\multicolumn{1}{c}{NGC} & \multicolumn{1}{c}{R$_{GC}$} &   \multicolumn{1}{c}{z} &
\multicolumn{1}{c}{$\Pi$} & \multicolumn{1}{c}{$\Theta$} & \multicolumn{1}{c}{W} \\
& \multicolumn{1}{c}{(kpc)} & \multicolumn{1}{c}{(kpc)} & \multicolumn{1}{c}{(km s$^{-1}$)} &
\multicolumn{1}{c}{(km s$^{-1}$)} & \multicolumn{1}{c}{(km s$^{-1}$)} \\
\tableline
\\
2808 & 11.0 & -1.9 & -82 (13) & 74 (18) & 70 (22) \\
3201 & 8.9 &   0.8 & -24 (12) & -301 (08) & 131 (09) \\
4372 & 7.0 & -1.0 &  39 (15) & 114 (14) & 77 (12) \\
4833 & 7.0 & -0.9 &  116 (20) & 22 (22) & -43 (11) \\
5927 & 4.5 & 0.6 & -16 (22) & 227 (17) & 35 (15) \\
5986 & 4.2 & 2.4 & 25 (27) & 13 (15) & 31 (19) \\  
\tableline 
\end{tabular}
\end{center}
\end{table}

Orbital parameters have been calculated as in Paper III. 
We have used the Johnston, Spergel \& Hernquist (1995, JSH95) potential model to integrate the orbits.
This model includes a bulge, disk and a spherical dark halo, and is widely used as
a tool to investigate orbits in a simple, analytical form for the Galactic potential.
The orbital parameters are averages over a 10 Gyr time interval.
Uncertainties were derived from the width of the distributions of orbital
 parameters over repeated integrations with
different initial positions and velocities. The estimate of the uncertainty is 
taken to be half of the interquartile range,
which is defined as the inner $50\%$ of the data.
Orbital integrations were repeated in a Monte Carlo 
fashion based on the uncertainties in the observed quantities:
proper motions, distance and radial velocity. 
Naturally, these derived orbital parameter 
uncertainties do not reflect uncertainties in the potential model and in the 
LSR properties. 

Results of the orbit integrations are presented in Tables 4 and 5, 
where the uncertainties are the values in parentheses.
The integrals of motion and
orbital periods are listed in Table 4, while the pericenter, apocenter radii, maximum 
distance from the plane, eccentricity and orbital inclinations are presented in Table 5.
We derive two orbital periods, the azimuthal one, and the
radial one. The radial period characterizes the interval between pericenter (apocenter) passages
and is smaller than the azimuthal one, because of the precession of the orbit. Uncertainties in the
radial period are similar to those  in the azimuthal one, therefore we have
not added them to column six of Table 4. We also include the total 
angular momentum L which is not a strictly conserved quantity for the JSH95 potential. However it
does provide some insight into the orbits since it can be thought of as a third integral of motion.
The value of L in Table 4 is the average over one orbital integration. 
%Two uncertainties are provided for this quantity. The value in brackets is derived from the
%intrinsic dispersion of L within one set of initial conditions (i.e., a single
%orbital integration), and it reflects the variation of L due to the
%non-spherical symmetry of the potential. 
As with the other orbital parameters, the uncertainty in $<$L$>$ is determined 
from multiple orbital integrations as the 
initial conditions are varied according to the uncertainties in the
measured quantities.

\begin{table}[tbh]
\begin{center}
\caption{Integrals of Motion and Orbital Periods}
\begin{tabular}{rrrrrr}
%\multicolumn{6}{c}{Table 4. Integrals of Motion and Orbital Periods} \\ \\
\tableline
\\
\multicolumn{1}{c}{NGC} & \multicolumn{1}{c}{E$_{orb}$} & \multicolumn{1}{c}{L$_z$} & \multicolumn{1}{c}{$<$L$>$} &
\multicolumn{1}{c}{P$_{\varphi}$} & \multicolumn{1}{c}{P$_{r}$} \\
 & \multicolumn{1}{c}{($10^4$ km$^2$s$^{-2}$)} & \multicolumn{1}{c}{(kpc~km~s$^{-1}$)} 
& \multicolumn{1}{c}{(kpc~km~s$^{-1}$)} & \multicolumn{2}{c}{($10^6$ yr)} \\
\tableline
\\
 2808   & -7.7 (0.3)  & 813 (103)      & 978 (105)  & 240 (14) & 154  \\
%        & -10.0 (0.2) & 813 (107)      & 946 & 255 (15) & 165 \\
 3201   & -4.3 (0.3)  & -2668 (079)    & 2891 (099)& 461 (26) & 315 \\
%        & -6.4 (0.3)  & -2668 (075)    & 2893 & 606 (39) & 428 \\
 4372   & -9.7 (0.2)  & 807 (041)      & 917 (054) & 156 (06) & 106 \\
%        & -11.9 (0.2) & 807 (039)      & 926  & 159 (07) & 106 \\
 4833   & -10.0 (0.3) & 150 (092)      & 463 (054)  & 154 (09) &  91 \\
%        & -12.2 (0.3) & 150 (094)      & 546  & 148 (08) &  96 \\
 5927   & -10.2 (0.2) & 1030 (027)     & 1063 (032) & 147 (05) &  99 \\
%        & -12.4 (0.2) & 1030 (027)     & 1065 & 138 (06) &  95 \\
 5986   & -11.9 (0.4) & 54 (068)       & 199 (044) & 107 (10) &  62 \\
%        & -14.0 (0.4) & 54 (064)       & 243  & 117 (09)  &  62 \\
\tableline 
\end{tabular}
\end{center}
\end{table}

\begin{table}[tbh]
\begin{center}
\caption{Orbital Parameters}
\begin{tabular}{rrrrrr}
%\multicolumn{6}{c}{Table 5. Orbital Parameters} \\ \\
\tableline
\\
\multicolumn{1}{c}{NGC} & \multicolumn{1}{c}{r$_{p}$} & \multicolumn{1}{c}{r$_{a}$} 
& \multicolumn{1}{c}{z$_{max}$} & \multicolumn{1}{c}{ecc.} & \multicolumn{1}{c}{$\Psi_{r}$} \\
%& \multicolumn{1}{c}{$\Psi_{L}$} \\
& \multicolumn{1}{c}{(kpc)} & \multicolumn{1}{c}{(kpc)} & \multicolumn{1}{c}{(kpc)} & &
\multicolumn{1}{c}{($\arcdeg$)} \\
% & \multicolumn{1}{c}{($\arcdeg$)} \\
\tableline
\\
2808 & 2.6 (0.4) & 12.3 (0.7) & 3.8 (0.3) & 0.65 (0.05) & 18 (1) \\
%& 34 \\
%     & 2.5 (0.4) & 12.6 (0.7) & 3.9 (1.0) & 0.67 (0.05) & 19 (4) & 31 \\
3201 & 9.0 (0.2) & 22.1 (1.4) & 5.1 (0.5) & 0.42 (0.02) & 18 (1)  \\
%& 23 \\
%     & 8.9 (0.2) & 28.2 (1.9) & 6.3 (0.5) & 0.52 (0.02) & 19 (1) & 23 \\
4372 & 2.8 (0.2) & 7.4 (0.2) & 1.6 (0.2) & 0.45 (0.04)  & 18 (2) \\
%& 28 \\
%     & 2.9 (0.3) & 7.3 (0.1) & 1.7 (0.2) & 0.43 (0.03)  & 20 (2)  & 29 \\
4833 & 0.7 (0.2) & 7.7 (0.7) & 1.8 (0.4) & 0.84 (0.03) & 20 (5) \\ 
%& 71 \\
%     & 1.0 (0.1) & 7.3 (0.8) & 2.2 (0.5) & 0.77 (0.02) & 28 (5)  & 74 \\
5927 & 4.5 (0.1) & 5.5 (0.3) & 0.7 (0.1) & 0.10 (0.03) & 9 (1) \\
%& 14 \\
%     & 4.5 (0.1) & 5.2 (0.3) & 0.7 (0.1) & 0.07 (0.03) & 9 (1)  & 15 \\
5986 & 0.6 (0.2) & 5.0 (0.5) & 1.9 (0.3) & 0.79 (0.04) & 29 (5) \\
%& 74 \\
%     & 0.9 (0.2) & 5.0 (0.5) & 2.4 (0.3) & 0.69 (0.04) & 28 (2)  & 77 \\
\tableline 
\end{tabular}
\end{center}
\end{table}

Improved potential models that more accurately describe the inner region of the Galaxy have been used
in other studies.
A recent example is the work by Allen et al. (2006) who find that orbits
do not differ considerably between a bar model and an axisymmetric one for clusters that 
do not reside within the bar region of our Galaxy. The cluster orbits that are
affected by the bar
tend to have larger radial and vertical excursions than in the axisymmetric case. 
For the clusters presented here, perhaps most prone to the bar potential is NGC 5986.
Since its orbit is already highly eccentric, the effect of the bar will not 
change the overall shape of the orbit.

\section{Discussion}

As inferred from the velocities, NGC 5927's orbital parameters confirm its membership
to a rotationally supported system, the thick disk. This is not surprising
considering its high metallicity (Armandroff 1988). 
The remaining clusters that have metallicities
consistent with membership to the halo (Zinn 1985), have orbits that 
generally confirm this membership.
These orbits have moderate to high orbital eccentricities, and a broad range 
in orbital angular momentum. It is somewhat intriguing that the orbital inclinations
for all of the five metal poor clusters
are rather low (Table 5), while the average value is $\Psi_{r} \sim 37\arcdeg$ for the entire
sample of 43 clusters with [Fe/H] $ < -1.0$ (Paper III and recent updates).

With the exception of NGC 2808 and 3201, all clusters spend their time within 
the Solar circle; however NGC 2808
does penetrate the inner Galaxy region. NGC 3201 is the most energetic cluster in the sample.
We have therefore checked whether its orbit projected onto the sky  matches the two recent
streams found in the SDSS: 1) the $63\arcdeg$-long, narrow stream reported by Grillmair \& Dionatos (2006), and
2) the ``orphan stream'' found by Belokurov et al. (2006) and Grillmair (2006). The orbit of NGC 3201
does not match either of the paths of these two streams. In fact, the maximum distance from the
Galactic plane reached by NGC 3201 (Table 5) is less than the current distance from the
plane of both of these two streams:
8 kpc for the $63\arcdeg$-long stream and 16 kpc for the orphan stream. 

Two recent papers (Frinchaboy et al. 2004, Martin et al. 2004) have suggested that a number of 
globular and open clusters may be associated  with
the ring-like Monoceros structure (SDSS, Newberg et al. 2002)
and the Canis Major overdensity (Martin et al. 2004). 
These suggestions were based on the Galactic location and radial velocity of the clusters.
Among these clusters, NGC 2808 was a candidate.
Our data however rule out this association given the very eccentric orbit of NGC 2808
(Table 5), and the 
thick-disk-like orbit of the Monoceros structure (Pe\~{n}arrubia et al. (2005). In fact,
none of the metal poor clusters in our sample can be associated with the Monoceros/Canis Major
 structures, on account of their highly eccentric or retrograde orbits.

NGC 2808 is a massive cluster with a well-documented extended blue HB
(e.g., Bedin et al. 2000, D'Antona et al. 
2005 and references therein). In fact, D'Antona et al. (2005) demonstrate that NGC 2808's main sequence has 
a spread blueward of the fiducial main sequence which can be explained by a He-enhanced population
($Y\sim 0.4$ for $20\%$ of the population). HB models with an enhanced He population also reproduce the
peculiar HB morphology of this cluster (D'Antona et al. 2005). This is the second case of a
globular cluster where the main sequence indicates He enhancement, the first one being the
remarkable $\omega$ Cen (Bedin et al. 2004, Norris 2004, Piotto et al. 2005), the most massive 
cluster of our Galaxy.
$\omega$ Cen also has a very extended blue HB that can be well explained in the framework
of He enhancement (Lee et al. 2005). Unlike $\omega$ Cen, very little to no metallicity spread is found in
NGC 2808 (Carretta et al. 2006). $\omega$ Cen is now widely believed to be 
the nucleus of a satellite galaxy captured and destroyed by the gravitational field of the
Milky Way, mainly on account of its  chemical abundance patterns that indicate
strong self-enrichment and multiple episodes of star formation (e.g., Smith 2004). 
More recently, a new picture has
emerged for all globular clusters with extended blue HBs. Since the He enhancement
appears to explain well the peculiar HB morphology (see also the case of NGC 6441, Calois \& D'Antona 2007),
and since most theoretical studies point to self-enrichment from a previous generation of 
massive stars in their AGB phase
as the source of the high He abundance (e.g., Karakas et al. 2006 and reference therein),
it has been suggested that all clusters with extended HBs may have been cores of 
disrupted dwarf galaxies (Lee et al. 2007). Lee et al. (2007) also show that these extended HB clusters
are the most massive in our Galaxy. Recent models of He enrichment 
from a previous generation of massive stars within a globular cluster-size system 
are however unable to reproduce the very high He abundance ($Y \sim 0.4$) inferred 
in NGC 2808 and $\omega$ Cen (e.g., Karakas et al. 2006, Bekki \& Norris 2006). 
This too has prompted the hypothesis
that such systems are born early on at the bottom of the potential well of a more massive system
($M \sim 10^7 - 10^8 M_{\odot}$, Bekki \& Norris 2006, Bekki 2006)
than what they currently retain; and subsequently the halos of these systems 
are disrupted and destroyed by the Galactic tidal field. 
None of the current models are however able to explain both the amount of He enrichment
and the particular abundance patterns seen in these globular clusters (the Na-O anticorrelation and the C+N+O
constancy for instance, Karakas et al. 2006, Bekki et al. 2007, Romano et al. 2007). 

Regardless of the difficulties
of models to reproduce in detail these abundance patterns, the suggestion that these clusters
may have originated in rather massive satellite systems has 
prompted a closer look at the properties of their
orbits. Thus, according to the newly emerged picture,
the orbits of the progenitors of these extended HB clusters should have been 
particularly prone to orbital decay, because they were massive and underwent
dynamical friction, and disruption, as they reached the denser, inner regions of our Galaxy.
 Models of the disruption of the host system of 
$\omega$ Cen by Tsuchiya et al. (2003, 2004) that aim to reproduce its present orbit and mass
indicate that the original system started with an orbital
eccentricity of 0.90 and an apocenter radius of 58 kpc, while the current values are $\sim 0.6$ and
$\sim 6$ kpc (Paper III, Allen et al. 2006). Along this line of reasoning, NGC 2808 
may be analogous. It has a relatively high
orbital eccentricity (0.65), and does not move farther than $\sim 12$ kpc from the Galactic center
(Table 5). In our sample, besides NGC 2808, cluster NGC 5986 may also belong to this category 
of systems with extended HBs (Alves et al. 2001, Rosenberg et al. 2000). It 
is also a rather massive cluster (Table 2). Its orbit is highly eccentric, practically 
plunging, and confined to within
the inner 5 kpc of the Galactic center.

A rather unexpected result from the orbits derived for the metal poor clusters is that
there are two pairs of clusters that have very similar orbital parameters.
The first pair consists of NGC 5986 and NGC 4833 (Tables 4 and 5), and the second
pairing is NGC 2808 and NGC 4372. Clearly, orbit angular momenta L$_z$, pericentric radii, eccentricities
and orbit inclinations agree very well (within $1\sigma$) for the pair NGC 5986 - NGC 4833,
and moderately well for the pair NGC 2808 - NGC 4372.
The most significant difference is the total orbital energy
difference between the two clusters in either pair, which implicitly affects
the apocenter radii, maximum distance from the Galactic plane and, to some extent the
eccentricity. The orbital energy difference between the two clusters of either pair
is $\Delta E_{orb} \sim 2 \times 10^4$ km$^2$ s$^{-2}$. For reference, we take the example
of Sgr and cluster Pal 12 which is now believed to have been torn
from Sgr. An initial argument supporting this picture was the analysis of their orbits
(Dinescu et al. 2000).  However, later on, other evidence strengthened this view:
the chemical abundance pattern of Pal 12 matches that of stars in Sgr (Sbordone et al. 2006, Cohen 2004),
and the cluster is embedded in Sgr tidal debris (Mart\'{i}nez-Delgado et al. 2002).

Assuming that indeed Pal 12 was torn from Sgr, we have calculated orbits
for these two systems using 
the proper motion from Dinescu et al. (2000) for Pal 12, and 
the proper motion of Sgr given by Dinescu et al. (2005), both in the JSH95 potential.
We obtain a difference between the orbital energy of Pal 12 and Sgr of $2.5 \times 10^4$ km$^2$ s$^{-2}$.
This simple-minded exercise leads to the suggestion that 
the clusters in each pair may be dynamically associated, and therefore had a
common origin in satellites of the size of Sgr. From the theoretical point of view,
disruption events of satellite galaxies described by Helmi \& de Zeeuw (2000) 
in the E$_{orb}$-L$_z$ and L-L$_z$ planes (their Fig. 4 for instance) show ranges 
in E$_{orb}$ and L compatible with the ranges of our two pairs of clusters.
For a given satellite, it is  L$_z$ that has the narrowest range in the Helmi \& de Zeeuw simulations,
and indeed these values agree within errors for our two pairs of clusters.
If indeed the clusters in each pair are dynamically associated, this reinforces
the hypothesis that each pair was born in a massive satellite system subsequently
destroyed.

Alternatively,
another simple-minded exercise is to estimate the chance of obtaining two apparent ``pairs'' of
dynamically associated clusters, drawn from a system which has velocities distributed randomly according to the
velocity ellipsoid of the halo. We have therefore assigned velocities drawn randomly
from a velocity ellipsoid with dispersions $(\sigma_{\Pi},\sigma_{Theta},\sigma_{W}) = (138, 104, 111)$ km s$^{-1}$
and averages equal to zero in each velocity component. The velocity dispersions are taken
from Paper III, for the metal-poor halo sample ([Fe/H] $< -0.8$). We have thus generated a set of
one hundred such random representations for the five metal-poor clusters in our sample. The integrals of motion were 
calculated for each generated representation, and then we have searched for ``pairs'' of clusters within
a given volume in the integrals-of-motion space. 
The search box in this space is based on the observed separations and
uncertainties of our actual measures for the two proposed cluster
pairs in our sample.
For example, the L$_z$ side of the box is calculated from
the quadrature sum of the difference between L$_z$ for one of our 
tentative cluster pairs and
the uncertainty in this difference as given by the values in parantheses in Table 4.
Thus, for the pair NGC 2808-NGC 4372 the search box is
($\Delta$L$_z$ , $\Delta$E, $\Delta$L) = $(111, 2.03 \times 10^4, 133)$,
and for NGC 5986-NGC 4833 is ($\Delta$L$_z$ , $\Delta$E, $\Delta$L) = $(149, 1.96 \times 10^4, 273)$
(units are those in Table 4). For the NGC 2808-NGC 4372 pair we obtain a $23\%$ chance of finding an apparent
dynamically associated pair from a system that has velocities randomly distributed
according to the known velocity ellipsoid of the halo, while for the
NGC 5986-NGC 4833 pair, we obtain a $30\%$ chance. Taken together, the chance 
that both of these cluster pairings are mere coincidence is $7\%$.

To this extent, we have shown that in a sample of five metal-poor clusters there appears to be 
clumping in the integrals-of-motion
space that is not likely due to chance occurence. This clumpiness should be further tested and
quantified by analyzing the entire sample of globular clusters with 3D velocities, and by comparing the data
with more realistic models of the formation of the  globular-cluster system such as those in Prieto \& Gnedin
(2007) for instance. We hope to be able to address this in a future paper.

This work was supported by NSF grants AST-0407292 and AST-0407293.
We thank the referee who has suggested the exercise concerning the 
chance occurence of pairs of dynamically-associated clusters.

\end{document}